\documentclass[
,final                                       
]
{aipproc}

\layoutstyle{8x11double}
\usepackage{amssymb}   
\usepackage{amsmath}
\usepackage{amsfonts}
\usepackage{amssymb}
\begin{document}

\title{
Unparticle Phenomenology -- A Mini Review}


\author{Kingman Cheung}{
  address={Department of Physics, National Tsing Hua University, Hsinchu 300, Taiwan},
  altaddress={Physics Division, National Center for Theoretical Sciences, Hsinchu 300, Taiwan}
}

\author{Wai-Yee Keung}{
  address={Department of Physics, University of Illinois, Chicago, IL 60607-7059, USA}
}

\author{Tzu-Chiang Yuan}{
address={Institute of Physics, Academia Sinica, Nankang, Taipei 11529, Taiwan}
}

\begin{abstract}
We review some collider phenomenology of unparticle physics,
including real emissions and virtual exchanges of unparticle.
Existing experimental constraints from 
collider physics as well as astrophysics are briefly discussed.
\end{abstract}

\maketitle

\section{Introduction}

\begin{figure}[t!]
\centering
\includegraphics[height= 6cm , width = 8cm]{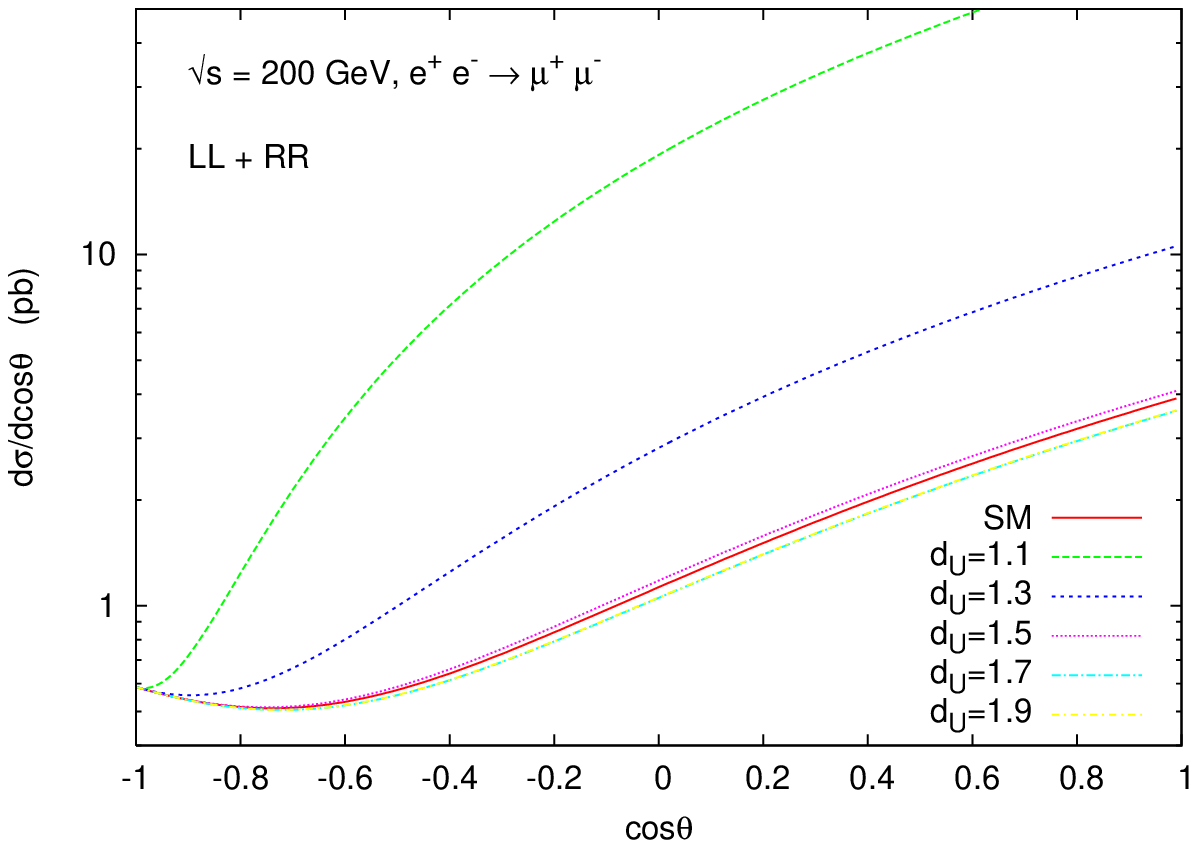}
\includegraphics[height= 6cm , width = 8cm]{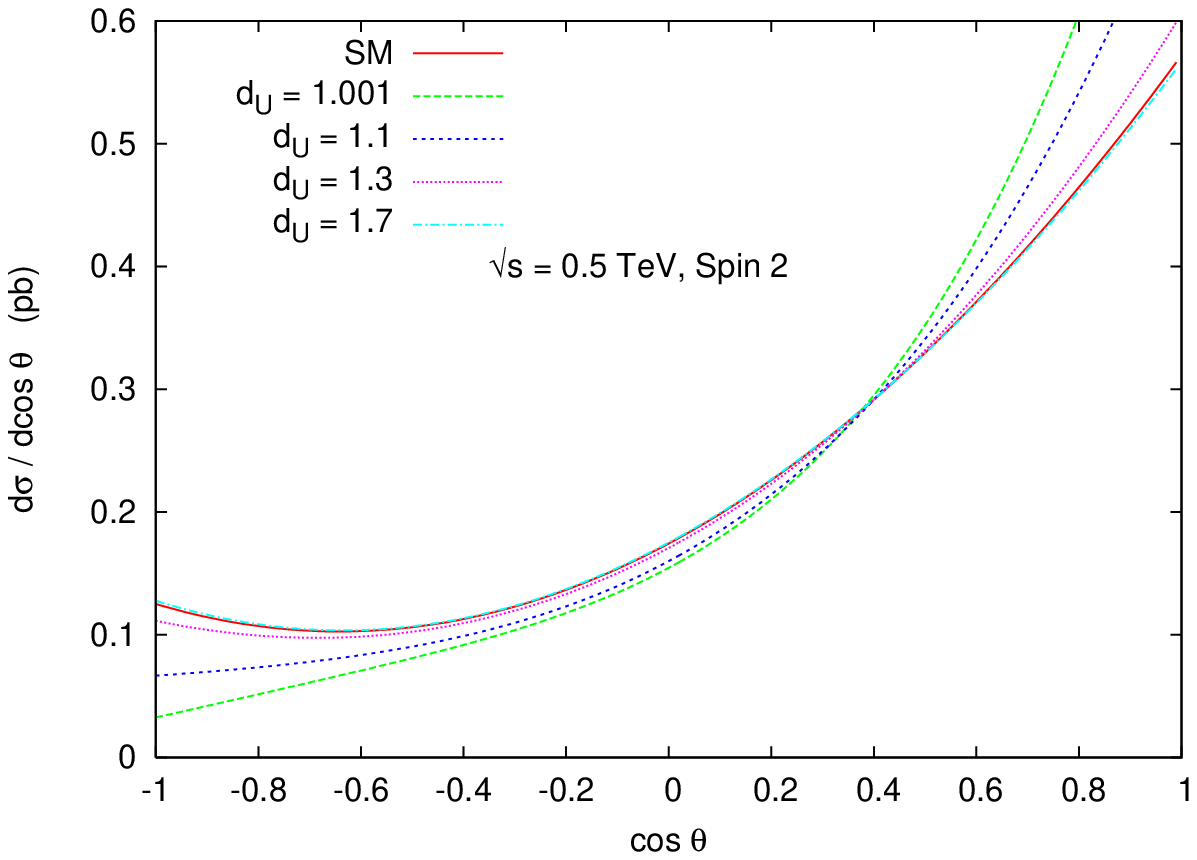}
\caption{\small \label{ee-ff-spin1-spin2}
Angular distributions for $e^-e^+ \to \mu^- \mu^+$ with various scaling dimension $d_U$ for the spin 1 unparticle exchange
with $LL+RR$ contact terms plus the SM contributions
at $\sqrt{s} = 200$ GeV  (left panel) and
for the spin 2 unparticle exchange plus SM contributions at $\sqrt s$ = 0.5 TeV (right panel). 
We have set $\Lambda_{U} = 1$ TeV and $\lambda_1 = \lambda_2 = 1$. See \cite{CKY-long} for details.
}
\end{figure}
The notion of unparticle introduced by Georgi \cite{unparticle} was based on the hypothesis that
there could be an exact scale invariant hidden sector resisted at a high energy scale. 
A prototype model of such sector is given by the Banks-Zaks theory \cite{Banks-Zaks}
which flows to an infrared fixed point at a lower energy 
scale $\Lambda_{U}$ through dimensional transmutation.
Below $\Lambda_{U}$, unparticle physics emerges and manifest itself as 
interpolating fields $O$ of various scaling dimensions and Lorentz structures. 

One of the interesting feature of unparticle operator 
is that it has a continuous spectral density $\rho(P^2)$ as a consequence of scale invariance \cite{unparticle}
\begin{equation}
\label{spectral}
\rho ( P^2) = A_{d_U} \theta(P^0) \theta(P^2) (P^2)^{d_U - 2}
\end{equation}
where $d_U$ is the scaling dimension of 
the unparticle operator with 4-momentum $P$ and
$A_{d_U}$ is a free normalization factor. Due to their similar kinematical exponent, Georgi  \cite{unparticle}
chose $A_{d_U}$ to be the prefactor of $d_U$ massless particle phase space 
\begin{equation}
d{\rm LIPS}_{d_U} = A_{d_U} \left( \left( p_1 + p_2 + \cdots + p_{d_U} \right)^2 \right)^{d_U - 2} 
\end{equation}
where $p_i^2  = 0 \, (i=1,...,d_U)$. Thus
\begin{equation}
\label{normalization}
A_{d_U} = \frac{16 \pi^{5/2}}{(2 \pi)^{2 d_U}} \frac{\Gamma( d_U + \frac{1}{2}) }{ \Gamma( d_U - 1) \Gamma( 2 d_U)} \; .
\end{equation}
As $d_U$ approaches 1, Eq.(\ref{spectral}) reduces to the familiar one massless particle phase space. 
This suggests unparticle behaves like a collection of $d_U$ massless particles. Since $d_U$ can be non-integral, 
one can now speak of fractional particle. This metaphor draws immediate attention of many physicists worldwide
as well as the general public.\footnote
{One can even find an entry of unparticle in the Wikipedia database.}

Soon after the introduction of unparticle, its propagator was deduced by using unitarity cuts \cite{georgi2} and 
the spectral decomposition formula \cite{CKY-short}. For a scalar unparticle, its Feynman propagator is given by
\begin{equation}
\label{propagator}
\Delta_F ( P ) = \frac{A_{d_U}}{2 \, \sin \left( d_U \pi \right)} \frac{1}{\left(  - P^2 - i \, 0^+ \right)^{2 - d_U}} \; .
\end{equation}
The presence of an extra CP-conserving phase factor 
$\exp \left( - i d_U \pi \right)$ in (\ref{propagator})
for non-integral $d_U$ and time-like $P^2 > 0$ 
was pointed out in \cite{georgi2, CKY-short} and 
the intricate interference effects in phenomenology
were also first studied there.
Analogous formulas can also be written down for the spin 1 \cite{georgi2, CKY-short} and 
spin 2 \cite{CKY-long} unparticle 
operators, while the fermionic spin 1/2 case was given 
in \cite{fermion-unparticle}. 

Unparticle operators can interact with the Standard Model (SM) fields via exchange
of some heavy particles of mass $M$. Integrating out such heavy fields induces a series of effective 
operators describing how unparticle interact with SM fields at low energy \cite{unparticle}.    
Some examples of these operators are \cite{unparticle,CKY-short}
\begin{eqnarray}
\mathrm{Spin \; 0:} &&
\lambda_0^\prime  \frac{1}{\Lambda_{U}^{d_U - 1}} \overline f f O\;, \;\; 
\lambda_0^{\prime\prime} \frac{1}{\Lambda_{U}^{d_U-1} } \overline f i \gamma^5 f O\;, \;\; \nonumber \\
&&
\lambda_0 \frac{1}{\Lambda_{U}^{d_U }} G_{\alpha\beta} G^{\alpha\beta}O \;, \; \cdots
\nonumber \\
\mathrm{Spin \; 1:} &&\lambda_1 \frac{1}{\Lambda_{U}^{d_U - 1} }\, \overline f \gamma_\mu f \,
O^\mu \;, \;\;
\lambda_1^\prime \frac{1}{\Lambda_{U}^{d_U - 1} }\, \overline f \gamma_\mu \gamma_5 f \,
O^\mu \;, \; \cdots
\nonumber \\
\mathrm{Spin \; 2:} && 
- \frac{1}{4}\lambda_2^\prime \frac{1}{\Lambda_{U}^{d_U} } \overline \psi \,i
   \left(  \gamma_\mu \stackrel{\leftrightarrow}{{\bf D}}_\nu + 
           \gamma_\nu \stackrel{\leftrightarrow}{{\bf D}}_\mu \right )
  \psi  \,  O^{\mu\nu} \; , \nonumber \\
  &&
  \lambda_2 \frac{1}{\Lambda_{U}^{d_U} } G_{\mu\alpha} 
G_{\nu}^{\;\alpha} O^{\mu\nu} \; , \; \cdots
\nonumber 
\end{eqnarray}
and so on. Here $O, O^\mu$ and $O^{\mu\nu}$ denote the spin 0, 1 and 2 unparticle 
operators respectively with $d_U$ its corresponding scaling 
dimension,\footnote{To simplify notation, we do not distinguish the  
various scaling dimensions of the unparticle operators of different spins.} 
$f$ is a SM fermion field, $G^{\alpha\beta}$ is a 
SM gauge field strength, $\bf D$ is a gauge covariant derivative acting on the SM fermion doublet $\psi$.
The $\lambda$-coefficients in front of these operators are effective couplings depend
on the short distance physics among the heavy exchange particles with the hidden sector and the 
SM sector. Thus they are free parameters and a priori of order unity. 
Besides the above operators, more complete lists of gauge invariant operators 
have been written down in \cite{Chen-He}. 
Antisymmetric rank 2 tensor unparticle operator was discussed in \cite{Hur-Ko-Wu}.

The hypothesis  of scale invariance of unparticle physics can be extended further to become 
conformal. The implication to unparticle physics of unitarity constraints on the scaling dimensions 
deduced from general conformal field theories \cite{Mack} was emphasized in \cite{Nakayama}.
The unparticle propagators for this general case have been worked out in \cite{GIR}.
Other theoretical aspects of unparticle were reviewed by Rajaraman \cite{Rajaraman-plenary}.

\vspace{-0.25cm}

\section{Indirect interference effects}

The complex nature of the unparticle propagator (\ref{propagator}) 
for non-integral $d_U$ and time-like $P^2$ can give rise to 
intricate interference patterns among the amplitudes of the unparticle and SM fields exchange 
in various elementary processes that are relevant in linear as well as hadronic colliders. 
Georgi \cite{georgi2} showed that the interference effects depend sensitively on the scaling dimension 
in the total cross section as well as in the forward-backward asymmetry of the process $e^- e^+ \to \mu^- \mu^+$. 
These effects are most easily palpable around the $Z$ pole. Similar 
effects were demonstrated for the Drell-Yan process at the Tevatron in \cite{CKY-short}.
More detailed studies of these two processes and some other related  ones
were presented in \cite{CKY-long}. 
Similar analysis of these complicated interference patterns 
due to spin 0 and spin 2 unparticle exchange
at the Large Hadron Collider (LHC) were given in \cite{Mathews-Ravindran-dilepton} for the Drell-Yan process and in
\cite{KMRT-diphoton} for the diphoton production. 
Analogous processes were also studied in \cite{CKY-long} for a future International Linear Collider (ILC). 

In Fig. \ref{ee-ff-spin1-spin2}, typical example of the 
sensitive dependence on the scaling dimension of a spin 1 and spin 2 unparticle operator 
in the intricate interference patterns is shown for the angular distribution  
of $e^- e^+ \to \mu^- \mu^+$  at the LEP2 energy and at 
the ILC energy of $\sqrt s = 0.5$ TeV respectively.

\begin{figure}[th!]
\centering
\includegraphics[clip,width=3.1in]{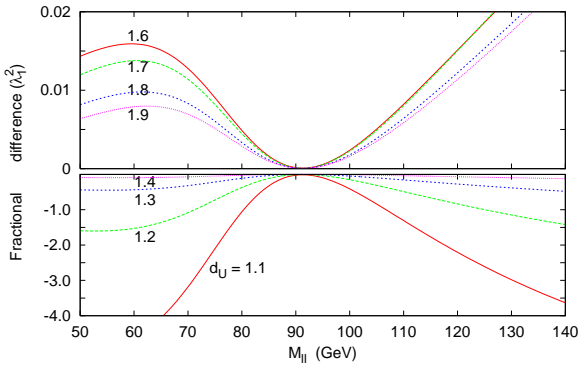}
\caption{
\label{drell-yan} \small
Fractional difference from the SM prediction of the Drell-Yan
invariant mass spectrum for various $d_U$ at the Tevatron in
units of $\lambda_1^2$.  We have chosen $\Lambda_U = 1$ TeV.  Note
that the scales in $\pm y$-axis are different. The curve for
$d_U=1.5$ is too close to zero for visibility in the current scale.
See \cite{CKY-short} for details.
}
\end{figure}

One of the most transparent effects of unparticle phase factor 
$\exp(-i d_U \pi)$ can be observed 
in the lepton-pair invariant mass spectrum near the $Z$ pole in Drell-Yan
production.
In Fig.~\ref{drell-yan}, we depict the 
fractional difference from the SM prediction 
in units of $\lambda_1^2$ (with small $\lambda_1$ 
while keeping $\Lambda_U = 1$ TeV) of the Drell-Yan distribution as
a function of the invariant mass of the lepton pair for various 
$d_U$. Interesting interference patterns around the $Z$ pole are easily 
discerned.

Three different groups \cite{photon-collider} 
have also studied these interference effects 
for the process $\gamma \gamma \to \gamma \gamma$ 
at a photon collider.

The static limits of unparticle propagators of various spins can also give rise to new long range forces that 
are severely constrained by low energy experiments. Coupling of these types of new forces 
to the energy-momentum stress tensor \cite{ungravity},  the baryon number current \cite{Desh-Hsu-Jiang}, 
the spin-spin interactions among electrons \cite{Liao-Liu} 
as well as the solar and reactor neutrinos \cite{Gonzalez-Garcia-etal,Anchordoqui-Goldberg}
have been studied in the literature. 

\vspace{-0.25cm}

\section{Direct production of unparticle stuff}

\begin{figure}[t!]
\centering
\includegraphics[width=3.2in]{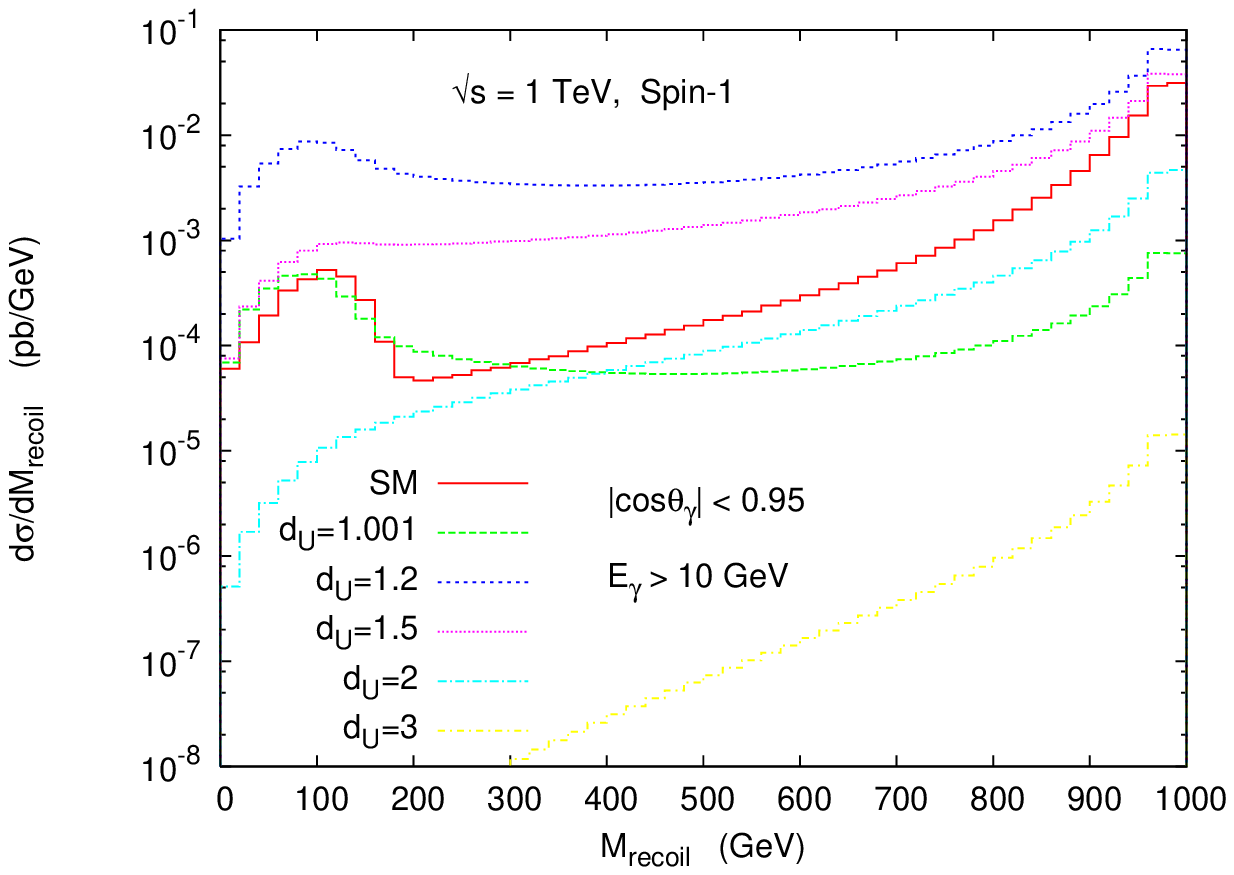}
\includegraphics[width=3.2in]{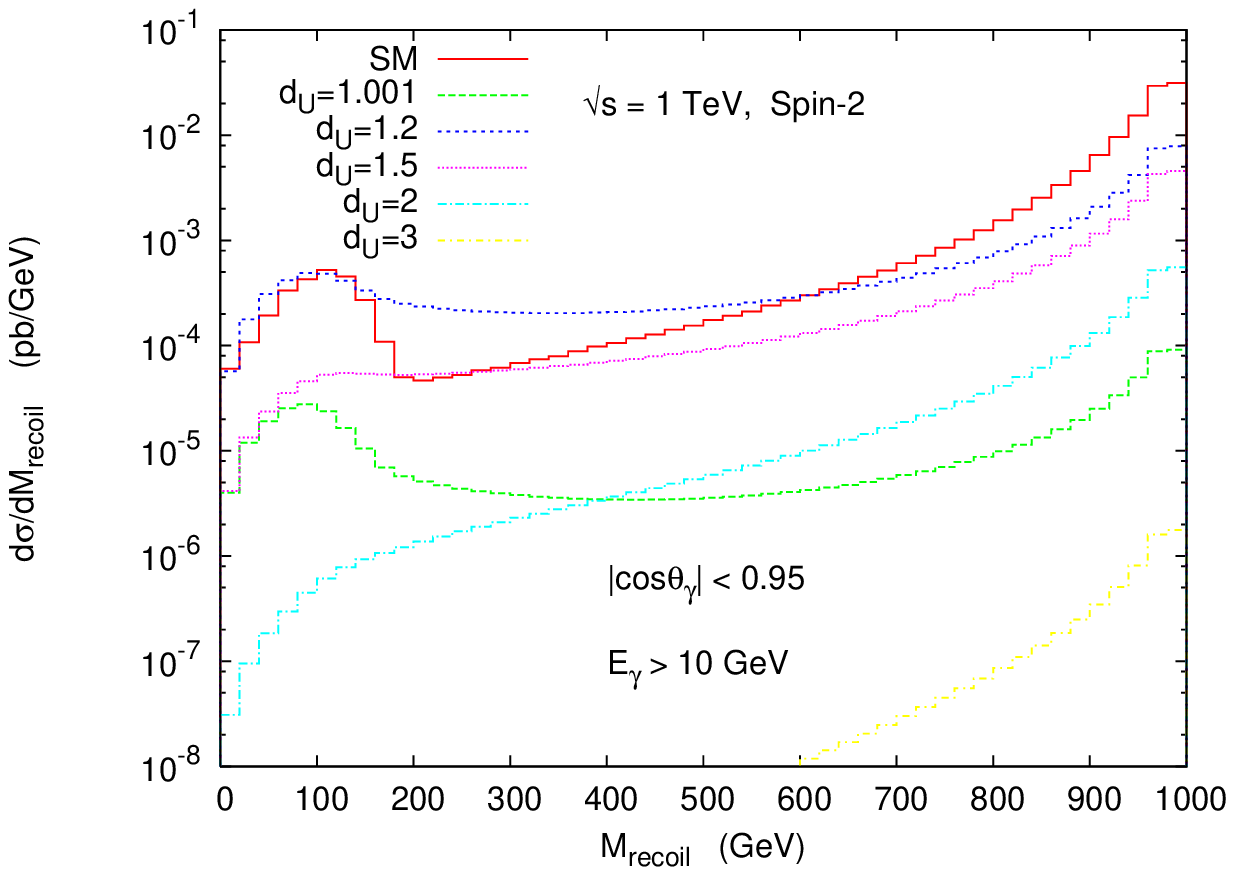}
\caption{
\label{eegammaU-spin1-spin2} \small
Comparison of recoil mass distributions 
of $e^- e^+ \to \gamma U$ with 
the SM background $e^-e^+\to \gamma Z^* \to \gamma \nu \bar \nu$ 
for different values of scaling dimension $d_{U}=1.001,\,1.2,\,1.5,\,2$ and 3 at 
the center-of-mass energy $\sqrt s = 1$ TeV. 
Left and right panels are for spin 1 and spin 2 unparticle respectively. 
See \cite{CKY-long} for details. 
}
\end{figure}
Unparticle operator $O$ with a continuous spectral distribution given by Eq.(\ref{spectral}) implies
the stuff it creates from the vacuum will have an indefinite dispersion relation given by 
$P^2 \equiv (P^0)^2 - {\bf{P}}^2 \ge 0$ with $P^0 \ge 0$. This stuff was coined as  unparticle \cite{unparticle}
since this is quite peculiar from the particle point of view that we so get used to.\footnote{
Recall that the wave-particle duality, a basic ingredient of quantum theory, is captured mathematically by a
definite dispersion relation.} 
We will denote this unparticle stuff generically by $U$. Signatures for detection of unparticle stuff are the missing 
energy and momentum carried away by the unparticle \cite{unparticle}. Search strategies for unparticle stuff are 
therefore similar to the search of Kaluza-Klein modes in the large extra dimension models. Even in a 2 body decay
of $1 \to 2 + U$, the kinematics of the particle 2 is no longer fixed due to the indefinite dispersion relation of $U$.

Many processes have been studied for real emission of unparticle stuff. 
These include $t \to b U$ \cite{unparticle}, $Z \to f \overline f U$ \cite{CKY-short,CKY-long}, 
Higgs$\to \gamma U $ \cite{CLY}, $Z \to \gamma U$ \cite{CKKY,Chen-He-Tsai}, 
$e^- e^+ \to (\gamma, Z) U$ \cite{CKY-short,CKY-long,Chen-He}, 
quarkonia$\to \gamma U$ \cite{Chen-He-Tsai} etc.  Similar to the indirect interference effects,
sensitive dependence on the scaling dimensions of the unparticle operators can be observed in the 
missing energy/momentum distributions for these processes. Production of multi-unparticle in the 
final state was considered in \cite{Feng-Rajaraman-Tu}.

Production of monojet plus unparticle was also studied in details at the LHC \cite{CKY-short,CKY-long}. 
It was shown that the strong dependence on the scaling dimensions 
of the matrix elements at the parton level 
is completely 
washed out at the hadron collider due to parton smearing effects. However, it has been demonstrated 
in \cite{Rizzo-monojet} that the monojet shape is still useful 
to distinguish the unparticle signals from large extra dimension models and the SM monojet background.

The mono-photon recoil mass distributions of $e^- e^+ \to \gamma U$ for spin 1 (left panel) 
and spin 2 (right panel)
unparticle $U$ are
plotted in Fig.~\ref{eegammaU-spin1-spin2} for various choices of $d_U$ 
at the center-of-mass energy $\sqrt s = 1$ TeV. 
The SM background from 
$e^- e^+ \to \gamma Z^* \to \gamma \nu \bar \nu$ is also displayed 
for comparison. The peculiar feature of unparticle effects due to the non-integral values of $d_U$
is easily seen in these recoil mass distributions. This is in sharp contrast with the case in the hadronic machines.

\vspace{-0.25cm}

\section{Existing constraints}

The scaling dimensions of unparticle operators and the effective couplings $\lambda$s that 
parameterize our ignorance of the underlying unparticle physics can be constrained by
existing collider experiments \cite{CKY-long} 
as well as observations from astrophysics and cosmology 
\cite{Davoudiasl,Hannestad-Raffelt-Wong,Kumar-Das,Freitas-Wyler,Dutta-Goyal}.

\begin{figure}
\centering
\includegraphics[height= 4.8cm , width = 7.25cm]{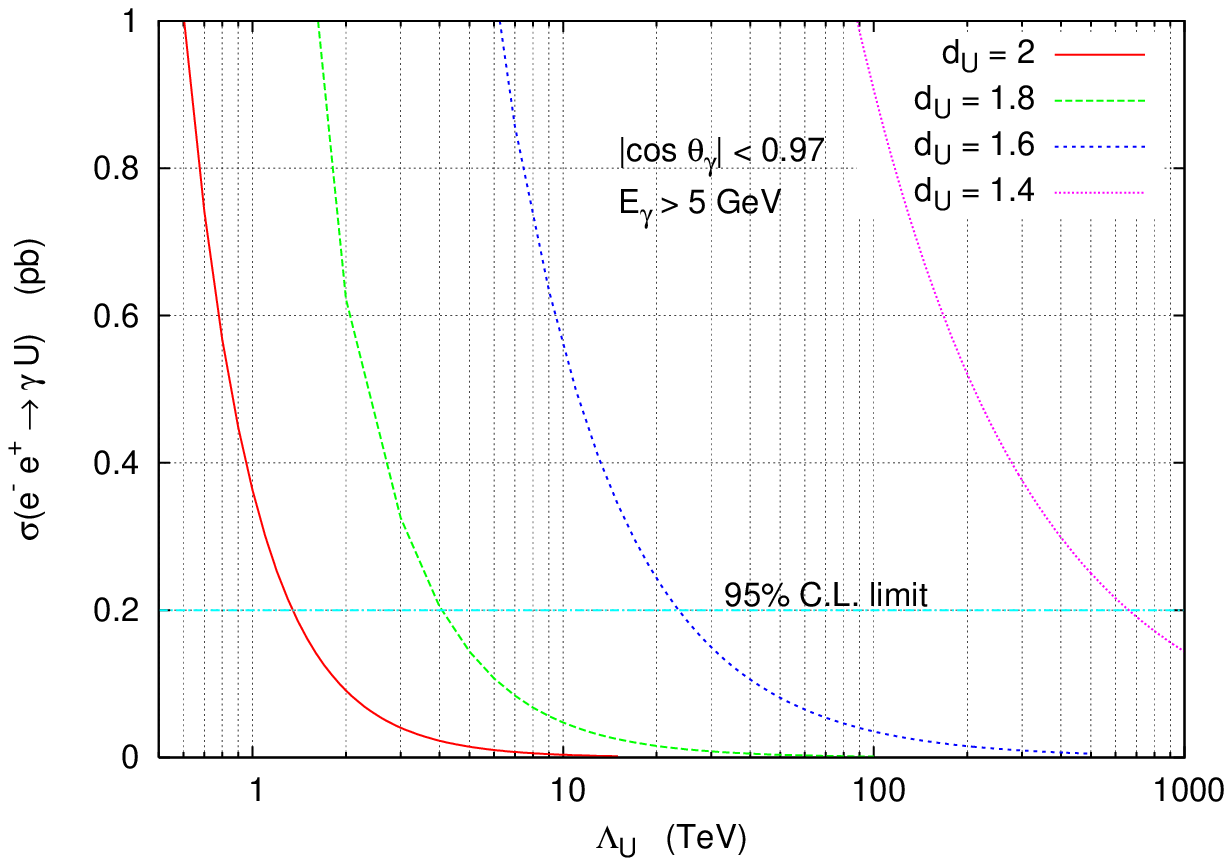}
\includegraphics[height= 4.8cm , width = 7.25cm]{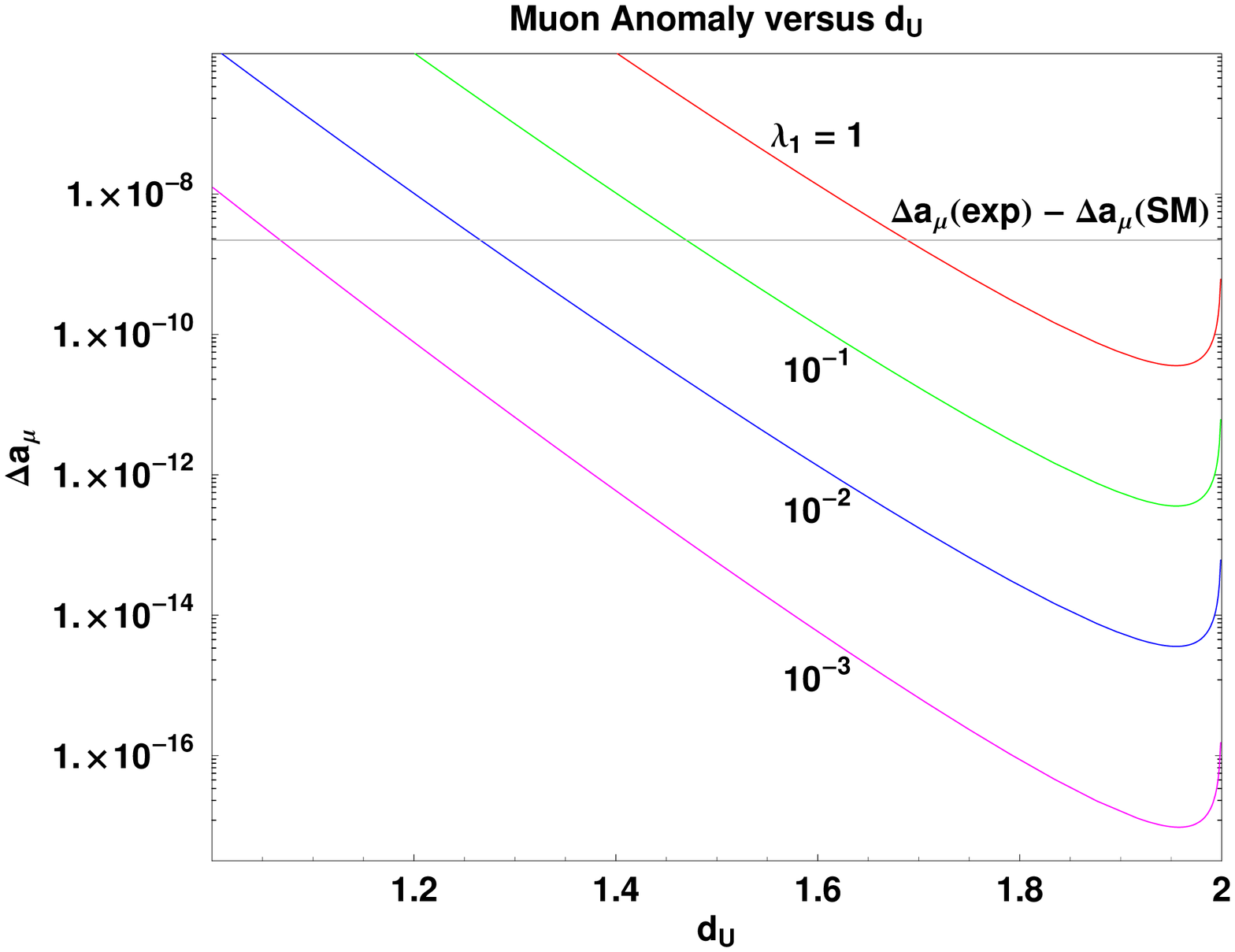}
\caption{\small \label{limits}
(Left panel) Cross sections for mono-photon plus unparticle production at 
the $e^- e^+$ collider with $\sqrt{s} = 207$ GeV for 
$d_U =$ 1.4, 1,6, 1.8 and  2. 
The horizontal line of 0.2 pb is the 95\% C.L. upper limit. (Ref. \cite{CKY-long})
(Right panel) Muon anomalous magnetic moment versus $d_U$ for 
$\lambda_1 = 10^{-n} (n=0,1,2,3)$. (Ref. \cite{CKY-short})
}
\end{figure}

Two constraints from different experiments are shown in Fig. \ref{limits}. 
In the left panel, the mono-photon plus missing energy production cross section measured 
at LEP2 \cite{lep-ph} was used to constrain $d_U$ and $\Lambda_U$. 
The limits are shown in Table \ref{table1}. 
\begin{table}[ht!]
\caption{\small \label{table1}
Limits on $\Lambda_U$ from mono-photon
production data of
$\sigma(e^- e^+ \to \gamma+X) \simeq 0.2$ pb at LEP2 (95\% C.L.)} 
\centering
\begin{tabular}{cc}
\hline
$d_U$   &   $\;\; \Lambda_U$ (TeV)  \\
\hline
$2.0$    &   1.35 \\
$1.8$    &   4   \\ 
$1.6$    &   23 \\
$1.4$    &   660 \\
\hline
\end{tabular}
\end{table}
In the right panel, constraints for $d_U$ and the effective vector coupling $\lambda_1$ 
can also be inferred from the muon anomalous magnetic moment. For the contribution from an antisymmetric 
rank 2 unparticle operator to the muon anomalous magnetic moment, see \cite{Hur-Ko-Wu}.

Assuming unparticle physics can extrapolate all the way down to the hot stars' interiors, 
emitting of such unparticle stuff can lead to star coolings. 
Observation of the cooling rates can thus place upper limits on the unknown parameters 
in the hidden sector of unparticle physics.
The cooling mechanisms may include many competing processes like
(1) un-Compton scattering $\gamma  e \to e  U$, 
(2) un-bremsstrahlung $ee \to eeU$, $eN \to eNU$ and $NN\to NNU$ where $N$ denotes a nucleus,
(3) photon-photon annihilation $\gamma\gamma \to U$, 
(4) electron-positron annihilation $e^-e^+ \to U$ and
(5) plasmon decay $\gamma^* \to \gamma  U$, etc. 
The calculation of the energy loss rate for each of these processes is similar to the axion case.

\begin{figure}
\centering
\includegraphics[height= 4.8cm , width = 7.25cm]{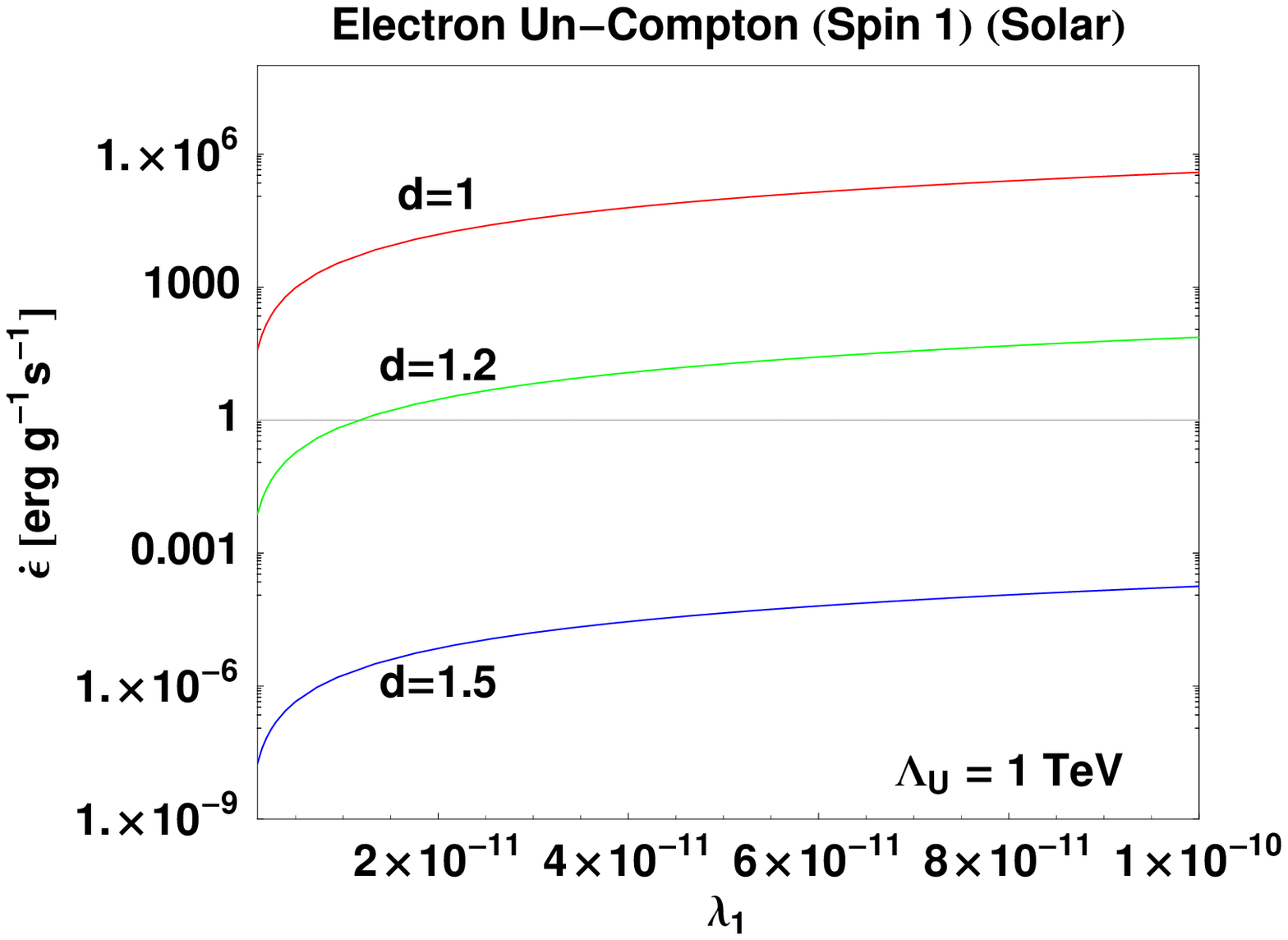}
\includegraphics[height= 4.8cm , width = 7.25cm]{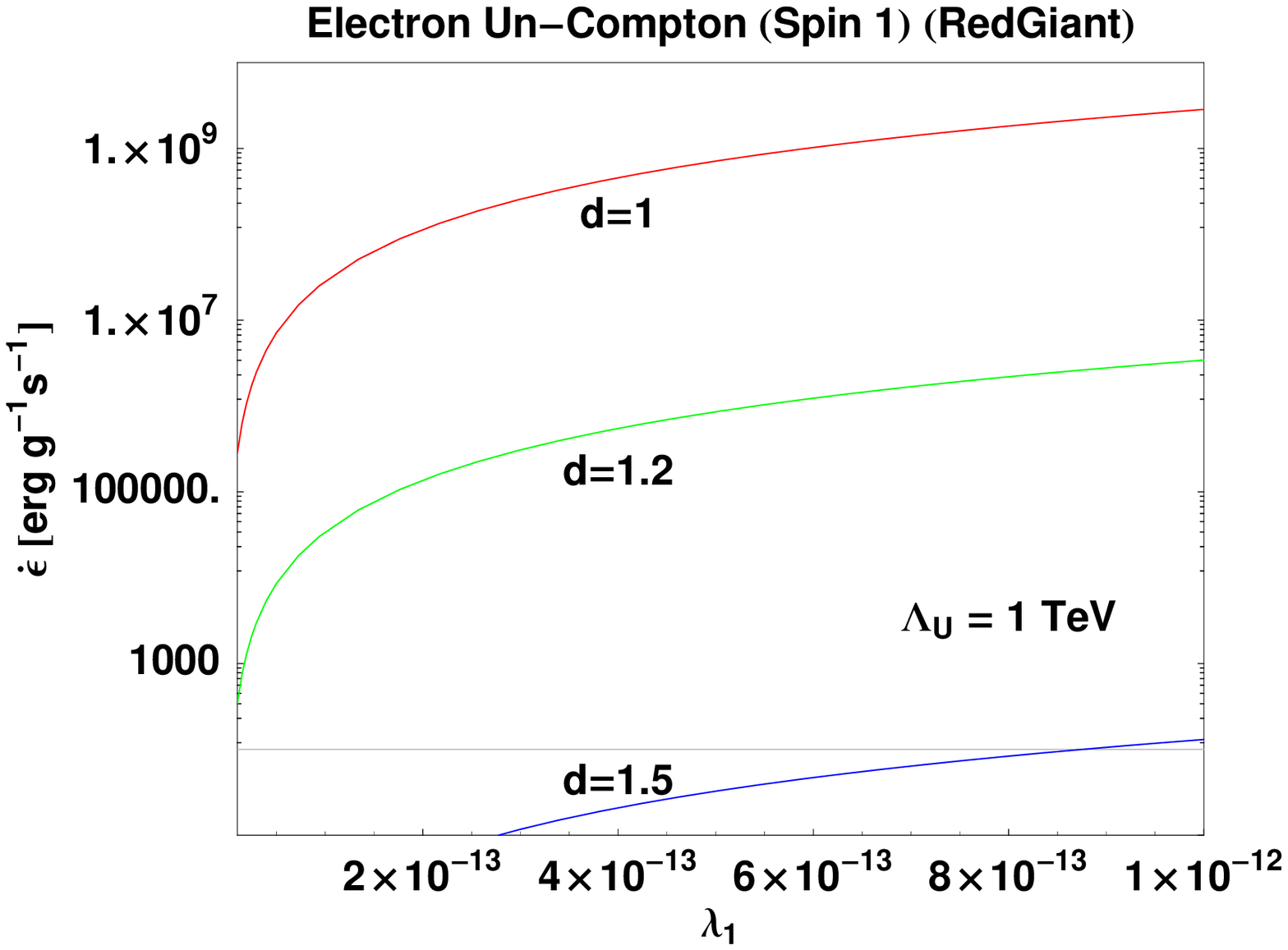}
\caption{\small \label{uncompton}
Energy loss rate $\dot \epsilon$ versus $\lambda_1$ for various values of $d_U$ due to the emission of 
spin 1 unparticle $U$ from the process $\gamma e \to e U$. 
The horizontal lines corresponds to 
$\dot \epsilon$(Solar) $\sim$ 1 erg g$^{-1}$ s$^{-1}$ (left panel) and  
$\dot \epsilon$(Red Giant)  $\sim$ 100 erg g$^{-1}$ s$^{-1}$ (right panel). \cite{CKY-unpub}
}
\end{figure}

As an example, we show in Fig. \ref{uncompton} the energy loss rate at the Sun and red giant 
from the emission of spin 1 
unparticle in the un-Compton scattering versus the effective coupling $\lambda_1$ for 
various values of $d_U$ \cite{CKY-unpub}. 
From these plots one can see that more stringent constraints on 
the hidden sector of unparticle physics 
can be deduced from astrophysics than from colliders. 
More detailed analysis of astrophysical constraints for the unparticle physics can be found in 
\cite{Freitas-Wyler}. Interesting constraints can be deduced from the null observation of any new long range force, 
and they were discussed in \cite{ungravity,Desh-Hsu-Jiang,Liao-Liu,Freitas-Wyler}. 
For constraints deduced from recent solar and reactor neutrinos data, 
see \cite{Gonzalez-Garcia-etal, Anchordoqui-Goldberg}.

\vspace{-0.25cm}
\section{Concluding Remarks}

We conclude with some remarks.
\begin{itemize}
\item
Scale invariance is expected not to be an exact symmetry at low energy.
Indeed coupling of a scalar unparticle operator with the SM Higgs field 
can break scale invariance once the Higgs field develops 
its vacuum expectation value \cite{Fox-Rajaraman-Shirman}.
To model the symmetry breaking effects, 
a finite mass gap $\mu$ was inserted in the unparticle spectral density by hand 
\cite{Fox-Rajaraman-Shirman}.
The unparticle propagator needs to be modified accordingly and it can
lead to unresonance behavior discussed in \cite{unresonance}. 
When the probing energy and the mass gap $\mu$ are proximate, 
more careful treatment is necessary in order to avoid the blow up of the amplitudes. 
Inclusion of the proper self-energy  \cite{BGKMS}
and the decay width of unparticle \cite{Rajaraman} in its propagator 
have been suggested in the literature.

\item
Besides the topics we reviewed briefly above, 
the roles of unparticle in flavour physics \cite{unparticle-flavorphysics,Zwicky},
top quark physics \cite{unparticle-topquark},
Higgs physics \cite{CLY,unparticle-Higgs}, electroweak symmetry breaking \cite{Lee}, 
unitarity in longitudinal $WW$ scatterings \cite{unparticle-WW},
supersymmetry \cite{Zhang-Li-Li,Desh-He-Jiang}, 
dark matter \cite{Desh-He-Jiang,Kikuchi-Okada}, 
gauged unparticle \cite{gauged-unparticle},
gauge couplings unification \cite{unparticle-unification},
hidden valley models \cite{Strassler},
deconstruction via AdS/CFT \cite{Stephanov} or Pad\'e approximants \cite{manuel},
walking technicolor \cite{unparticle-walking}, 
baryon number violating nucleon decay \cite{He-Pakvasa},
uncosmology \cite{Davoudiasl,uncosmology},
conformal energy and charge correlations \cite{conformal-collider}
etc have also been explored in the literature.
This list is really a tall order for unparticle physics given 
the fact that its inception was solely based on curiosity.
Thus, we refer the readers to these references for details
and apologize to those authors whose works we do not mention here
due to page limitation.

\item
In less than two years, numerous unparticle phenomenological studies have been 
performed despite the physical meanings of many aspects of the unparticle are not yet clear.
How does dimensional transmutation occur that leads to the unparticle phenomena?
Is unparticle stable? What is the partition function of
a system of unparticle? How does one couple gauge field to unparticle? etc. These are interesting
theoretical questions awaiting for answers.

\item
Finally, we note that phenomenology of unparticle has been regarded as a little bit premature \cite{premature}.
We will let our readers to make their own judgements.

\end{itemize}

\newpage

\vspace{.5cm}
\begin{theacknowledgments}
This work was supported in part by the NSC under grant No. NSC 96-2628-M-007-002-MY3, 
the NCTS of Taiwan as well as U.S. DOE under grant No. DE-FG02-84ER40173.
\end{theacknowledgments}


\end{document}